# Trigger design for a gamma ray detector of HIRFL-ETF

DU Zhong-Wei(杜中伟) [1,2], SU Hong(苏弘) [2], QIAN Yi(千奕)[2], KONG Jie(孔洁)[2]

[1] *State Key Laboratory of Particle Detection & Electronics, University of Science and technology of China, Hefei 230026, China*

[2] *Institute of Modern Physics, Chinese Academy of Sciences, Lanzhou 730000, China*

E-mail: dzw@mail.ustc.edu.cn

**Abstract** The Gamma Ray Array Detector (GRAD) is one subsystem of HIRFL-ETF (the External Target Facility (ETF) of the Heavy Ion Research Facility in Lanzhou (HIRFL)). It is capable of measuring the energy of gamma-rays with 1024 CsI scintillators in in-beam nuclear experiments. The GRAD trigger should select the valid events and reject the data from the scintillators which are not hit by the gamma-ray. The GRAD trigger has been developed based on the Field Programmable Gate Array (FPGAs) and PXI interface. It makes prompt trigger decisions to select valid events by processing the hit signals from the 1024 CsI scintillators. According to the physical requirements, the GRAD trigger module supplies 12-bit trigger information for the global trigger system of ETF and supplies a trigger signal for data acquisition (DAQ) system of GRAD. In addition, the GRAD trigger generates trigger data that are packed and transmitted to the host computer via PXI bus to be saved for off-line analysis. The trigger processing is implemented in the front-end electronics of GRAD and one FPGA of the GRAD trigger module. The logic of PXI transmission and reconfiguration is implemented in another FPGA of the GRAD trigger module. During the gamma-ray experiments, the GRAD trigger performs reliably and efficiently. The function of GRAD trigger is capable of satisfying the physical requirements.

## 1 Introduction

HIRFL-ETF (The External Target Facility (ETF) of the Heavy Ion Research Facility in Lanzhou (HIRFL[1])) has been built for in-beam nuclear experiments. The heavy ion beam is accelerated from 10 MeV/u to 1.1 GeV/u in the main ring (CSRm) and sent into the ETF via a radioactive beam line (RIBLL II). The ETF experiments include the study of the nuclear matter (Uranium ions of above 500 MeV/u is used to collide the uranium target) and the study of properties and architecture of hypernuclei (high-energy proton is used to collide the normal nuclei to generate Λ hypernuclei). The design goals of the ETF electronic system are particle discrimination, momentum resolution and neutral particle detection with high precisions. The Gamma Ray Array Detector (GRAD), one subsystem of ETF, is based on CsI detectors and is capable of measuring the energy of gamma-rays with CsI scintillators. The GRAD is composed of an inner ring and an adjacent outer ring, each with 512 CsI scintillators, totally comprising 22% of a sphere. The 1024 scintillators generate a vast amount of data for the data acquisition (DAQ) system of GRAD. Most of the acquired data are useless including the data of invalid events and the data of scintillators that are not hit in collision events. So the GRAD trigger is developed for selecting the valid event data from detectors and saved trigger data for analysis. The background noise data that are not of interest to the experiment should be rejected efficiently in order to decrease the dead time of the DAQ system.

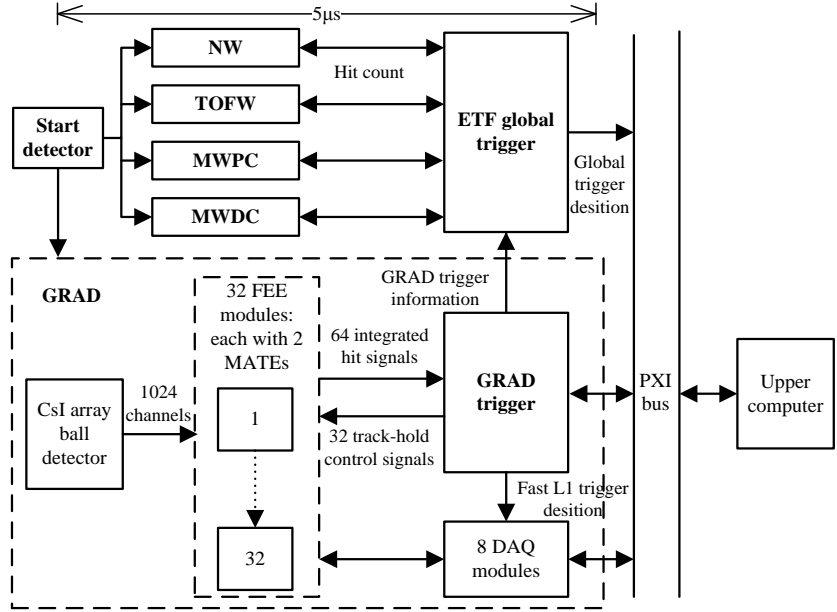

Fig. 1 The signal flow in the ETF trigger and GRAD electronics.

The CsI array ball of GRAD is divided into 64 detector sections, each with 16 CsI scintillators. The front-end electronics (FEE) is composed of 32 modules, each with 2 onboard application specific integrated circuit (ASIC) chips named MATE[2]. Each MATE chip is connected to a detector section, and it integrates the signals from the 16 CsI scintillators of the detector section into one fast hit signal via OR operation. A total of 64 fast hit signals (32 signals are related to the inner ring and 32 signals are related to the outer ring) are delivered to the GRAD trigger module, as shown in Fig. 1. In addition, the GRAD trigger module supplies 32 external control signals for the FEE in order to increase the time accuracy of track-and-hold operations in the MATE chips.

The ETF global trigger collects the preprocessed Level-1 trigger information from GRAD, neutron wall (NW), time-of-flight wall (TOFW), multiwire proportional chamber (MWPC) and multiwire drift chamber(MWDC). Because of the latency differences of these ETF sub-detectors, the ETF global trigger should wait for all the L1 trigger information for about 5 μs to make a global trigger decision. However, after receiving the hit signal from one scintillator, the FEE of GRAD will shape the signal and stop accepting new signals until a trigger signal arrives. That means a dead time is generated based on the latency of the trigger signal. In order to decrease the dead time of the GRAD electronics, the GRAD trigger is required to generate a fast trigger signal for the DAQ modules to decide whether to start read-out first and reset the FEE. Then the slow global trigger decision will decide whether the event data acquired by the DAQ modules should be stored.

The gamma-ray in-beam experiments have several features about trigger. The noise is of high level and of high frequency. In most collision events only one gamma-ray hits the GRAD detector. Therefore, in order to reach high trigger efficiency, the physics experiments dictate several general design requirements for the GRAD trigger. First, the hit signal whose amplitude is below the threshold should be rejected. Second, the hit signals of one valid event should be inside a window time. Third, several trigger conditions should be applied in selecting the valid events based on the physical requirements.

Concurrently, the GRAD trigger module is required to process a maximum of 64 fast hit signals at a speed of 100MHz and provide preprocessed GRAD trigger information for the GTS, consisting of hit number and condition information. The GRAD trigger module is required to deliver the valid physics event data to the host computer for off-line analysis.

Referring to the trigger systems of several particle experiment electronics, such as BESIII[3], ATLAS[4] and ALICE[5], the GRAD trigger satisfying the physical requirements has been developed. In this paper the design and the performance of the GRAD trigger are described in detail below.

## 2 The implementation of the GRAD trigger

According to the trigger requirements, the GRAD trigger method consists of three parts. First, the hit signals of 1024 scintillators are preprocessed with amplitude discriminating by 32 FEE modules to reject the ground noise. Second, the hit signals are aligned in a window time to discriminate different events. Third, the trigger algorithm based on five trigger conditions is implemented in the GRAD trigger module.

To have a reconfigurable GRAD trigger implemented in a single PXI module has been kept as a design principle. As is shown in Fig. 2, the GRAD trigger is implemented in a PXI-6U module containing 2 ALTERA FPGAs (EP1C12F324). The trigger processing logic and algorithm are all implemented in a kernel trigger FPGA. And the other FPGA implements PXI data transmission and reconfiguration for the kernel trigger FPGA. The GRAD trigger module performs four tasks: the hit-signal preprocessing, the kernel trigger processing, the data transmission via PXI bus and the reconfiguration of the trigger logic when the firmware needs upgrade.

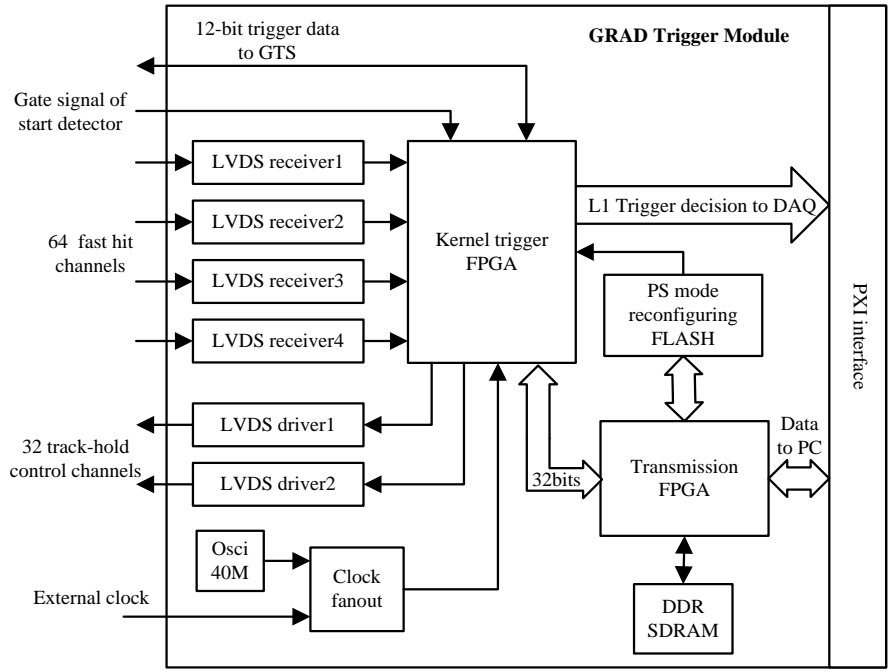

Fig. 2 Architecture of the GRAD trigger module.

## 2.1 Preprocessing

The preprocessing includes signal amplitude discrimination, synchronizing signals to the local clock and aligning the data among channels.

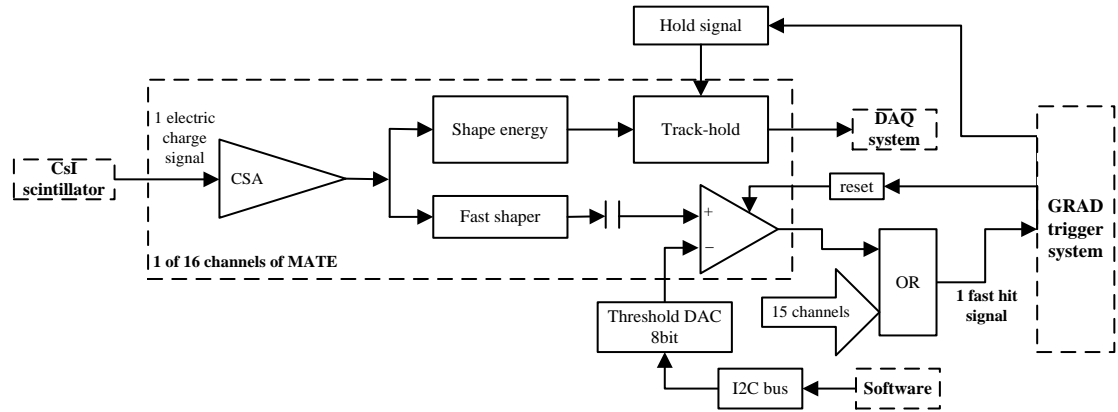

Fig. 3 The signal amplitude discriminating circuit in MATE

Fig. 3 shows the amplitude discrimination implemented in the MATE chip of the FEE. An electric charge signal generated by a CsI scintillator is converted to a voltage signal via a charge sensitive amplifier with fast shaper. An in-chip 8-bit digital-to-analog converter (DAC) is employed to generate the threshold voltage. The amplitude of the threshold voltage can be adjusted by configuring the DAC with the software based on the noise level. The signal is compared with the threshold voltage in order to reject the signals generated by the ground noise. After amplitude discrimination, the signals from 16 scintillators of the same detector section are converted to one fast hit signal with OR operation in one MATE chip. The fast hit signal is a Low Voltage Differential Signaling (LVDS) signal. A total of 64 LVDS hit signals (the FEE contains 64 MATE chips, each generates a hit signal) are delivered to the GRAD trigger module via twisted-pairs. Four 16-channel LVDS receivers (SN65LVDS386) are employed to convert the LVDS signals to Low Voltage TTL (LVTTL) signals in the trigger module.

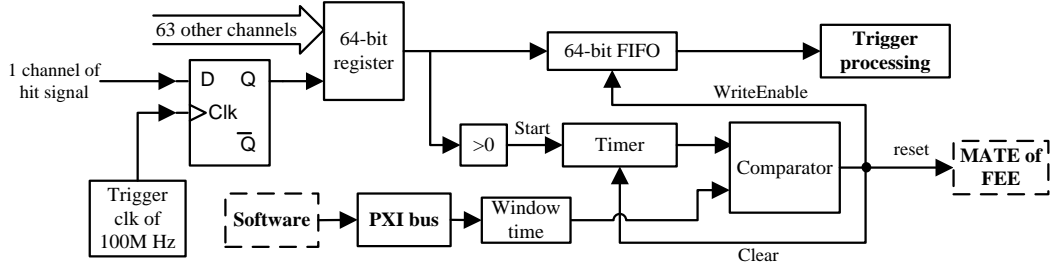

Fig. 4 The synchronizing and aligning circuit in the trigger FPGA

Then the 64 hit signals are received as a 64-bit trigger data by the kernel trigger FPGA. Each 1-bit data represents one hit signal from one detector section and should be set to "1" if a hit occurs on the corresponding 16 CsI scintillators. D flip-flops, a 64-bit register and a 64-bit first-in first-out buffer (FIFO) are employed to synchronize the hit signals with a trigger clock of 100 MHz. The clock is generated by an in-chip phase locking loop, and it is used for the whole trigger processing.

Due to the difference of hit position, the arrival time of charged particles that strike the corresponding detector section of each signal channel is different [6]. It is necessary to align all the channel data into one event for further trigger processing. The aligning process is shown in Fig. 4. When an event occurs, the input of the FIFO will be delayed for a window time. The window time is the maximum time difference between the arrivals of hit signals of all channels. The value of the window time depends on the physical condition and should be set via the software before the experiment. Due to the working mechanism of the MATE chip (if the MATE chip receives no reset signal from the trigger module, the hit signal will maintain "1" whatever the scintillator signals change), the trigger data will not be missed when the input of the FIFO is disabled in the window time. When the input of the FIFO is enabled, the data entering the FIFO will be the aligned data because the signals of all hit channels in the event have arrived already. Then the input of the FIFO will be disabled again after completing storing a 64-bit data of this event. A reset signal will be sent to the MATE chip. The preprocessing circuits will be ready for the next event and the aligned data will be delivered to the kernel trigger processing.

## 2.2 The kernel trigger processing

The kernel trigger processing of yielding trigger data and decision is implemented in the trigger FPGA. Because of the complexity of the algorithm, it is impossible to complete the trigger processing in a single clock cycle. Therefore, the hit data are processed in three clock cycles in the pipeline. In the first clock cycle, the hit numbers of the total detector are calculated by adding up the 64 bits of the hit data. The 64-bit data will be sent to the upper computer. So the software can delete useless data acquired by the DAQ modules based on the trigger data.

In the second clock cycle, the hit number and the hit data are processed to generate 5-bit trigger information based on five trigger conditions. The content of the 5-bit trigger information is listed in Table 1. The first and second trigger conditions are basic multiplicity trigger. However, in the experiments, the noise is of high level and its amplitude is close to the amplitude of hit signals. So only some of the noise is rejected by the signal amplitude discrimination. On the other side, in most collision events, only one gamma-ray is accepted by GRAD and only two or three adjacent channels are hit. That enables us to set only the multiplicity parameter as 2. Therefore, in order to further reduce the false trigger rate, the last three trigger conditions are applied. Compared with it under the trigger condition when the hit number $\geqslant$2, the false trigger frequency can be reduced to 1/32 on the trigger condition that there should be two adjacent channels hit in a valid event.

In fact, we choose some of the five trigger conditions based on the physical experiment in order to reach high trigger efficiency and low event missing rate. So in the third clock cycle, the 5-bit trigger information is compared with a 5-bit condition parameter. The condition parameter is set via the software which allows the user to configure the trigger parameters of GRAD trigger module. Each bit of the condition parameter should be set to "1" if the related trigger condition is enabled for making the trigger decision. If each bit of the trigger information is equal to or greater than each bit of the condition parameter, the physics event is judged to be valid. The 12-bit L1 trigger data composed of the 7-bit hit number and the 5-bit trigger information will be delivered to the GTS. Because the external clocks of all ETF electronics are generated by the same clock module, the transmission between the GRAD trigger module and the GTS is a simple serial process synchronously with the external clock of 40 MHz. Concurrently, a trigger decision will be sent to the DAQ system to start read-out of the event data. The 64-bit hit data will be stored in a 32-bit asynchronous FIFO and wait for delivery to the transmission FPGA. If the 5-bit trigger information is not equal to the 5-bit condition parameter, the physics event is judged to be invalid. The trigger logic and the MATE chips will be reset.

Table 1. Content of the 5-bit GRAD trigger information

| Index | GRAD Trigger Condition | Description |
|---|---|---|
| 1 | Total hit number$\geqslant$1 | The total number of hit channels is larger than or equal to 1 |
| 2 | Total hit number$\geqslant$N | The total number of hit channels is larger than or equal to N. N is set via the software. |
| 3 | Inter ring to inter ring hit | The gamma-ray goes though scintillators of two adjacent sections in the inter ring |
| 4 | Outer ring to outer ring hit | The gamma-ray goes though scintillators of two adjacent sections in the outer ring |
| 5 | Inter ring to outer ring hit | The gamma-ray goes though scintillators of the inter ring section and scintillators of the adjacent section in the outer ring |

## 2.3 PXI transmission and reconfiguration

Because the experiment environment is closed and of high radiation, the GRAD trigger module should be capable of upgrading the trigger logic remotely. So the transmission FPGA containing the logic of reconfiguring the kernel trigger FPGA is employed in the GRAD trigger module. Fig. 5 shows the logic of the transmission FPGA. When the trigger logic should be upgraded, a PS configuration file of the new trigger logic will be sent to an onboard serial flash memory (M25P80). After receiving reconfiguration order from the software, the PS controller will supply the PS time sequence for the configuration pins of the trigger FPGA. So the trigger FPGA will be reconfigured with the file read from the flash memory.

There are 32 dual-port data lines and 4 control lines between the two FPGAs. The transmission controller of the transmission FPGA uses the Wr/Rd line to switch among receiving trigger data and sending parameters via the data lines. The other 3 control lines are used to control read-out of the asynchronous FIFO in the trigger FPGA. The trigger data will be stored in a 32-bit FIFO and delivered to the host computer via the PXI bus.

The software in the host computer is developed by using Visual C++ and WINDRIVE which make the software design easier. A graphical user interface allows configuring parameters and options of the trigger logic and the transmission logic of the trigger module. The data received by the software will be used for off-line analysis and calibrating the data from the DAQ system.

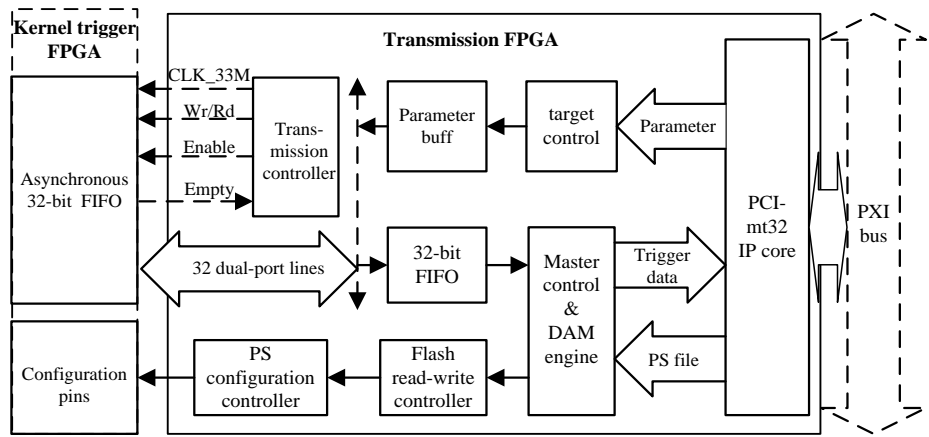

Fig. 5 Logic of transmission and configuration

## 2.4 FEE track-and-hold controlling

The track-and-hold operation in the MATE needs an external hold signal to keep the in-chip capacitor at the voltage peak. In fact, the time of signal reaching the peak depends on the shape of the input signal and the noise. So the GRAD trigger module is required to supply a user-adjustable hold signal for each FEE module. The hit signals delivered to the GRAD trigger module are fast shaped by the MATE chips. There are 32 timers in the trigger FPGA logic. After receiving either hit signal of one FEE module, the corresponding timer will start counting. When the timer time reaches the presupposed time, a hold signal will be generated. At the beginning of each experiment the time of signal reaching peak should be measured first to configure the time parameter. Two 16-channel LVDS drivers (SN65LVDS387) are employed to convert the LVTTL signals of the FPGA to the LVDS signals. The LVDS hold signals will be delivered to the corresponding FEE modules via twisted-pairs.

## 3 Performance of the GRAD trigger

The GRAD electronics has been assembled. Fig. 6 shows the photograph of the GRAD trigger module.

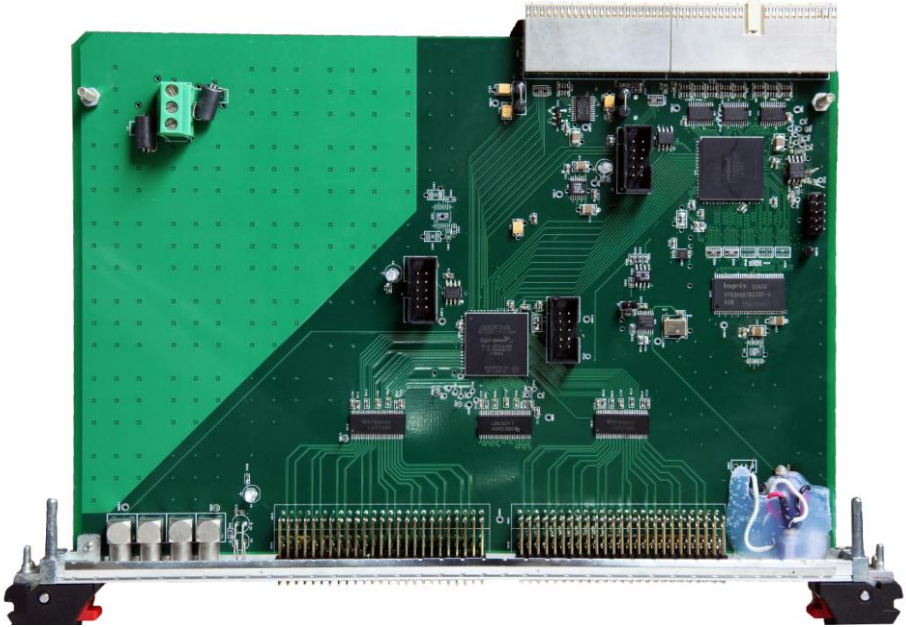

Fig. 6 (color online) The GRAD trigger module

The performance of the GRAD trigger module has been tested in the gamma-ray experiments together with 512 CsI scintillators of the GRAD, 16 FEE modules, and the DAQ system. In order to increase the trigger efficiency, signals of the start detector are used as the start time of valid events. The latency of trigger processing is less than 50ns and it is capable of meeting the experiment requirements. Table 2 shows the event frequency in several trigger conditions in one gamma-ray experiment. The frequency under GRAD trigger conditions matches the anticipated gamma-ray density and accords with the mechanism of the hit process shown in Fig. 7. When a gamma-ray hits a scintillator of the inter ring, it will be absorbed by the scintillators on the path. If the 16 scintillators of the hit section (link to one MATE chip and generate one hit signal) are not able to absorb all energy of the gamma-ray, the adjacent section (in the inter ring or in the outer ring) will also be hit. In fact, due to the direction of the gamma-ray, the adjacent section in the outer ring will have a higher possibility to be hit than the adjacent section in the inter ring. So the frequency of the inter-to-outer hit should be higher than that of the inter-to-inter hit and higher than that of the outer to outer hit, which accords with the test data. The test results demonstrate the reliable trigger functionality of the GRAD trigger.

On the other side, the noise of each channel is random and shows no correlation with it of other channels. So in the test system with 32 channels working, under Trigger Condition 5 the false trigger frequency caused by noise can be reduced to 1/16 of that under Trigger Condition 2. In fact, Table.2 shows that the trigger frequency under Condition 5 is 0.4% less than that under Condition 2. So we can come to the conclusion that under Trigger Condition 2 the false trigger rate is less than 0.5% and the trigger efficiency is more than 99.5%. The trigger efficiency that is close to 100% satisfies the physical requirement.

Actually, in experiments we should not just consider the trigger efficiency, but also consider the frequency of real events rejected under the trigger conditions. Therefore, we chose to enable Trigger Condition 2 in the experiments though the false trigger frequencies under the last three trigger conditions are even less. On the other side, gamma-rays of full-energy peak that are absorbed completely by one detector channel will be ignored under Condition 2.

Table 2. Event frequency of trigger conditions in the gamma-ray experiment

| Index | GRAD Trigger Condition | Frequency(Hz) |
|---|---|---|
| 1 | Total hit number≥1 | 729 |
| 2 | Total hit number≥2 | 692 |
| 3 | Inter ring to inter ring hit | 136 |
| 4 | Outer ring to outer ring hit | 158 |
| 5 | Inter ring to outer ring hit | 689 |

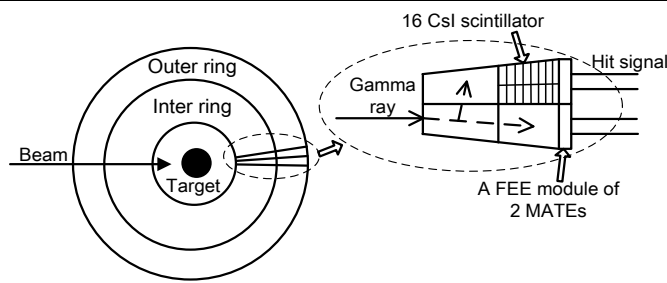

Fig. 7 The process of the gamma-ray hitting the detector

## 4 Conclusion

The GRAD trigger described in this paper has a reliable and efficient performance on yielding correct trigger information and decisions for the valid events. The gamma-ray experiments demonstrate that the GRAD trigger is of low latency and of high trigger efficiency. It is capable of satisfying the physical requirements. The reconfiguration capability of the GRAD trigger offers the advantage of flexible firmware upgrade. Due to the acceptable performance, the GRAD trigger module will be copied to apply in the building silicon strip detector in the HIRFL-ETF.

*This work was supported by the Important Direction Project of the CAS Knowledge Innovation Program (No.KJCX2-YW-N27) and the National Natural Science Foundation of China (No. 11005135).*